# DIVERSIFICATION AND HYBRIDIZATION IN FIRM KNOWLEDGE BASES IN NANOTECHNOLOGIES[1]

Avenel, E.♣; Favier, A.V., Ma, S.; Mangematin, V♥., Rieu, C♦.

The paper investigates the linkages between the characteristics of technologies and the structure of a firms' knowledge base. Nanotechnologies have been defined as converging technologies that operate at the nanoscale, and which require integration to fulfill their economic promises. Based on a worldwide database of nanofirms, the paper analyses the degree of convergence and the convergence mechanisms within firms. It argues that the degree of convergence in a firm's nano-knowledge base is relatively independent from the size of the firm's nano-knowledge base. However, while firms with small nano-knowledge bases tend to exploit convergence in each of their patents/publications, firms with large nano-knowledge bases tend to separate their nanoR&D activities in the different established fields and achieve diversity through the juxtaposition of the output of these independent activities.

[1] The paper is based on a research study carried out within the European Nanodistrict project, which is part of the Network of Excellence PRIME. We gratefully acknowledge EC funding. The paper benefits from stimulating discussions with our colleagues Ph. Laredo, A. Rip, D. Robinson, B. Kahane, M. Zitt, R. Stankiewicz, A. Bonaccorsi, C. Palmberg and T. Grid. We gratefully acknowledge the comments from Pr. B. Bozeman and two anonymous referees. Usual caveats apply.

♣ Corresponding author : eric.avenel@upmf-grenoble.fr
♥ Grenoble Ecole Management (GEM) France and UMR GAEL (INRA-Univ of Grenoble), BP 47X, 38040 Grenoble Cedex, France
♦ UMR GAEL (INRA-Univ of Grenoble), BP 47X, 38040 Grenoble Cedex, France

1. Introduction

The past 5 years have seen an explosion of interest in the area of science and technology labeled "nanotechnology". Although at an early stage, promises have lead to high expectations of the fruits that could be harvested from investment in nanotechnology development (Saxl, 2005). But how do firms develop nanotechnologies? Do they develop new independent fields of research or do they integrate nanotechnologies within their on-going research projects? In other words, do nanotechnologies develop within firms by juxtaposition of new R&D projects *independent* to the existing ones or do they develop by hybridization of nanotechnologies *within* existing projects?

Darby *et al.* (2003) suggest that the development of nanotechnologies is a *Grilichesian breakthrough* which follows a similar pattern to that of biotechnology. Based on Hill and Rothaermel (2003), they predict a relative decline of the economic performance of incumbents as a result of the emergence of this competence-destroying technology (Shea, 2005). However, nanotechnologies borrow not only from biotechnology but also from microelectronics. Abernathy and Utterback (1978) have underlined the critical role of large incumbents (such as Fairchild semiconductors, IBM and Texas Instruments) in the early development of micro-electronics during the 60s and 70s. Microelectronics and biotechnologies have followed two different evolutionary paths over recent decades. Predicting the type of path that nanotechnologies will follow is a difficult issue. Our data on firms performing nano-R&D show that both incumbents and new firms are investing in the development of nano-knowledge bases (NKB). The paper addresses the respective roles of incumbents and new firms focusing on *how* firms with different profiles develop their nano-knowledge bases.

One key dimension of this issue is that the field of nanotechnology covers multiple scientific disciplines and technological domains. Different reports (Rocco, 2002; Nordmann, 2004) emphasize that the realization of the potential of nanotechnologies is based on the convergence of technologies from physics, engineering, molecular biology and chemistry. This convergence may however be an artifact of the agglomeration of the scientific and/or technological output of a large number of heterogeneous players. Do the nano-knowledge bases (NKB) of individual firms also exhibit significant convergence amongst technologies? Does the degree of convergence depend on the size of firms' NKB? Do firms with NKBs of different size achieve convergence through similar strategies?

The paper formulates hypotheses which are tested on a worldwide database of nanotechnology firms ([www.nanodistrict.com](www.nanodistrict.com)). We identify several trajectories for the development of nano-knowledge bases by firms. Firms with small NKBs as well as those with large ones exhibit high degrees of convergence for scientific and technological fields. However, they obtain similar degrees of convergence through different arrangements. Large NKBs are collections of items focusing each on one technological/scientific field (collection of articles or patents in different fields) while in small NKBs each item is related to several fields (one or two articles or patents, each of them is related to several fields). This confirms the view that nanotechnologies emerge from the convergence of established fields and suggests that small firms have a greater ability to exploit the opportunities created by this convergence.

**2. Hybridization and the diversity of the nano-knowledge base**

The paper assesses firms' scientific knowledge base through publications and their technological knowledge base through patents. Convergence at the firm level is measured by the *diversity* of scientific/technological fields to which the portfolio of publications/patents is related. The diversity of a portfolio of items can be achieved in two different ways: juxtaposition and hybridization. We define *juxtaposition* as the collection of independent scientific and technological fields within the same NKB. This is typically what can be observed in firms performing independent R&D programs, each program being strongly embedded in one traditional field like physics or chemistry. Alternatively we define *hybridization* as the case in which each item is related to various fields: in effect, hybridization is diversity at the level of individual items. This is typically what will be observed in firms where programs are performed by teams grouping together researchers and engineers from widely different backgrounds.

Juxtaposition means that a firm cannot integrate nanotechnologies unless it enlarges its knowledge base, which might entail developing new labs, hiring new researchers, forming new alliances or even investing in new locations. On the contrary, hybridization means that nanotechnology competences and devices are integrated within existing research projects. New competencies are clearly linked to the existing ones and the size of the knowledge base remains stable. The ways by which nanotechnologies are developing within a firm will impact the diversity of its knowledge (*i.e.* the breadth of the firm knowledge base) and its research performance. Zhang *et al.* (2006) argue that firms with more targeted knowledge base are more efficient to perform research in the short run and they are more able to form alliances as they build a larger absorptive capacity. Porac *et al.* (2004) argue in a similar way when they analyze the impact of the heterogeneity of human resources within research teams. This discussion of hybridization vs. juxtaposition leads us to formulate two conflicting hypotheses.

*Hypothesis 1: When firms develop their nano-knowledge bases through hybridization, the diversity of the nano- knowledge base remains stable when its size increases.*

*Hypothesis 2: When firms develop their nano-knowledge bases through juxtaposition, the diversity of the nano-knowledge base increases when the size of the nano-knowledge base increases.*

Hypothesis 2 means that the presence of different technologies in the aggregate output of nanoR&D activities does not reflect the existence of a convergence at the level of firms' R&D programs and teams. Scientists and engineers with different backgrounds lead their research separately and their output is focused on separate fields. Conversely, hypothesis 1 implies that convergence is not an artifact of aggregation but rather a major characteristic of nanotechnologies that is reflected in the composition of research and development teams.

## 3. Characterizing the knowledge base of nanofirms

The nano S&T publications examined include all papers related to nanotechnologies indexed in the Thomson-ISI 'Web of Science' database between 1993 and 2003: around 122,000 publications have been identified. The nano-publications data has been obtained through the use of sophisticated scientometric methods (Zitt *et al.*, 2006) which have improved the basic bibliometric method (Meyer *et al.*, 2001). The scientometric methodology is a two-stage method based on keywords, which are used to identify and download all publications related to nano S&T. The first extraction made through keywords is controlled using citation method to control for the relevance and centrality of the publications to nano S&T. This process identified 1271 firms as publishing about 15,000 nano S&T articles over the 1993-2003 period. Corporate production of nano S&T publications was both sustained and increased over the period, with a significant difference between pre-1999 and post-1999 years.

The data on the number of nano-patents and research fields was obtained from USPTO (US

Patent and Trademark office) patents information. The extraction (which included the use of the TAG[2] nano defined *ex post*) formed a sample of 4,000 nano-patents in the 1993-2003 period. The second stage of data collection on patents was to identify all the patents filed by firms which filed at least one nano-patent. Nano-patents are very rare until the late 80s, though after 1989, there was a clear 'take-off', since when there has been an impressive growth in the number of nano-patents.

*Size.* Within the nano-knowledge base of firms, we distinguish between the scientific and the technological NKB. To measure the *size of the scientific nano knowledge base*, we take the decimal logarithm of the number of nano-publications of the firm. The *size of the technological nano knowledge base* is measured by the decimal logarithm of the number of nano-patents granted to the firm.

*Diversity of scientific and technological nano-knowledge base.* The diversity variables measure the breadth of a firm's nanoR&D activities to ascertain it they are concentrated in a small number or spread over a larger number of fields. Field definition was based on the ISI Journal Classification system for publications, and the US Patent Classification for patents. Borrowing a tool used in industrial organization to measure market concentration, we take 1 minus the Herfindahl index as our measure of diversity. This diversity index theoretically yields values between 0 and 1, with larger index values corresponding to greater diversity, but in practice no values over 0.8 were obtained in our sample.

---

[2] The TAG nano has been defined by patent offices to identify patent related to nanotechnologies.

**Please insert table 1 here**

*Hybridization of scientific and technological nano knowledge base.* To measure the extend to which firms exploit the opportunities that nanotechnologies create to organize the convergence of different fields within their research projects, the hybridization variables count the number of different technological/scientific fields quoted on average by each patent or publication of a specific firm. The variables are real numbers equal to or larger than 1, with larger values indicating higher degrees of hybridization.

Table 1 provides measures and summary data for each variable. Variables related to publications are calculated only for firms with at least one nano-publication, and those related to patents only for firms with at least one nano-patent.

## 5. Results

Figure 1 represents the evolution of diversity when the size of the technological nano-knowledge base of the firm increases. Hypothesis 2 suggests an increasing relation between the two variables, but this is not observed in figure 1. Indeed, there are very few firms below the 45-degree line, which means that firms with a large NKB also have a diverse NKB. This tends to support the juxtaposition hypothesis rather than the hybridization hypothesis. As the NKB of firms is developed through juxtaposition, its diversity increases with its size. However, an unexpected finding is that there is a quite significant set of firms above and on the left of the 45 degree line, that is, firms with a small but diverse knowledge base. This suggests that nano S&T is also developing through hybridization, as the diversity within nano S&T remains stable when the size of firm's nano knowledge base increases. This tends to support the hybridization hypothesis rather than the juxtaposition hypothesis. So, if a firm has a large knowledge base, it will be more diversified, but it is not true that if a firm is diversified it will necessarily have a large knowledge base. We find small firms which are diversified, and the same level of diversity is achieved by firms which differ a lot as regards the size of their knowledge base: there are clearly several possible firm profiles in the nano industry. Figure 2 displays a similar pattern to Figure 1, representing the diversity of the scientific nano-knowledge base compared to its size. As far as hypotheses 1 and 2 are concerned, we cannot confirm or reject either.

Two patterns of nanotechnology development within firms co-exist, juxtaposition of nanotechnologies to existing projects - in which diversity is linked to size - and hybridization, in which it is not. While it is easy to figure out a mechanism generating the first pattern, such as the juxtaposition of unrelated nanoprojects, the second one, where the expansion of the knowledge base is realized through hybridization, appears to be more difficult to explain.

**Please insert figures 1 and 2 about here**

As nanotechnologies have been defined as converging technologies at the crossroad of different scientific and technological fields, how to internalize nano S&T diversity remains a central question for firms. To examine further the issue as to how firms of different size achieve the integration of nanotechnologies, we divide our sample by the size of the nano-knowledge base and plot the values of the firms' hybridization index and diversity index. This leads us to figures 3 (patents) and 4 (publications). (We provide separated plots for various ranges of NKB size.

**Please insert figure 3 about here**

**Please insert figure 4 about here**

Figures 3 and 4 show that, while firms with large knowledge bases exhibit limited degrees of hybridization (and thus cluster on the top left of the figure), some small firms reach very high degrees of hybridization, which makes it possible for them to build diversified knowledge bases based from a limited number of patents or publications. New entrants in nanotechnologies - whether high tech start-ups or firms moving into the field - are those which are integrating nanotechnologies through hybridization, while firms which are already performing research in one of the technologies which form nanotechnologies (chemistry, microelectronics, biotechnologies etc.) develop new programs of research which focus on nanotechnology fields which are new to them. It seems reasonable to conclude that what is emerging now in nanoS&T neither reflect the development patterns of biotechnology or of microelectronics, but is a mixture of these two patterns, in which there is space for both start-ups and incumbents in the R&D activity.

## 6. Conclusion and Discussions

Our findings confirm the idea that different scientific/technological fields are converging in nanotechnologies. We find that firms' nano-knowledge bases are quite diversified, regardless of their size. It also turns out that firms are following quite different trajectories in the development of their nano-knowledge bases. Small firms, at least some of them, are achieving very significant levels of diversity through intense hybridization. Big firms, with a few exceptions, also have developed diversified nano-knowledge bases, but their use of hybridization is much more limited. The diversity of their portfolio is the result of the juxtaposition of items focused mainly on one or two established scientific/technological fields. This suggests that small firms are in a better position than big firms to exploit the opportunities created by the emergence of nanotechnologies. However, this conclusion would be too simplistic. In fact, the relative success of both incumbents (big firms) and new entrants (small firms) will be determined by several other elements. We discuss briefly two of these elements here. First, this paper does not directly address the issue of whether nanotechnologies are competence-destroying or enhancing. This would require us to look at the relation between firms' nano-knowledge base and their global knowledge base. Do the competences that incumbents build on to develop their nano-knowledge bases correspond to their existing non-nano-knowledge base? How does this issue relate to whether such firms develop their nano-knowledge base through juxtaposition? These are avenues for future research. Second, access to existing research and production facilities is a key asset in nano S&T. Thus access to large facilities such as those in Minatec or Albany can be particularly helpful for nanoelectronics developments, and to the facilities developed in universities and firms in the case of nanobiotechnologies. Such research facilities and technological platforms can be seen as specialized complementary assets which can improve incumbent performance when a radically new technology is introduced (Rothaermel *et al.*, 2005). Even if nanotechnologies are competence-destroying, incumbents may limit their risk through their superior access to these resources, just as pharmaceutical groups managed to cope with the emergence of biotechnologies because of their specific assets in the administrative validation and distribution of drugs.

**Table 1: Summary of Variables and Measures**

| | Variable | Measure | N | Mean | Median | Q1 | Q3 |
|---|---|---|---|---|---|---|---|
| Technological Nano-Knowledge Base | Size of Technological NKB | Decimal logarithm of number of nanopatents | 1003 | 0.28 | 0 | 0 | 0.48 |
| | Diversity of Technological NKB | 1-Herfindal index calculated in US Patent Classification (4-digit) fields | 1003 | 0.51 | 0.57 | 0.44 | 0.75 |
| | Hybridization in Technological NKB | Average number of IPC (4 digits) fields per patent | 1003 | 2.08 | 2.00 | 1.20 | 2.60 |
| Scientific Nano-Knowledge Base | Size of Scientific NKB | Decimal logarithm of number of nanopublications | 1271 | 0.55 | 0.48 | 0 | 0.90 |
| | Diversity of Scientific NKB | 1-herfindal index calculated in ISI Web of Knowledge subject categories | 1271 | 0.53 | 0.67 | 0.37 | 0.80 |
| | Hybridization in Scientific NKB | Average number of ISI subject categories per publication | 1271 | 1.52 | 1.41 | 1.00 | 1.86 |

**Figure 1: diversity and size of technological nano-knowledge base**

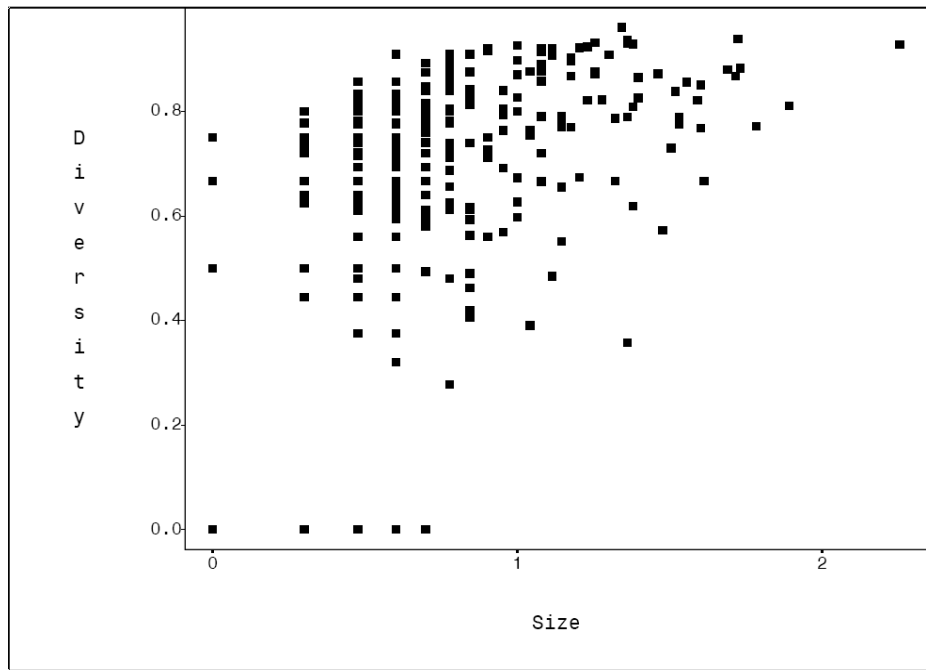

**Figure 2: diversity and size of scientific nano-knowledge base**

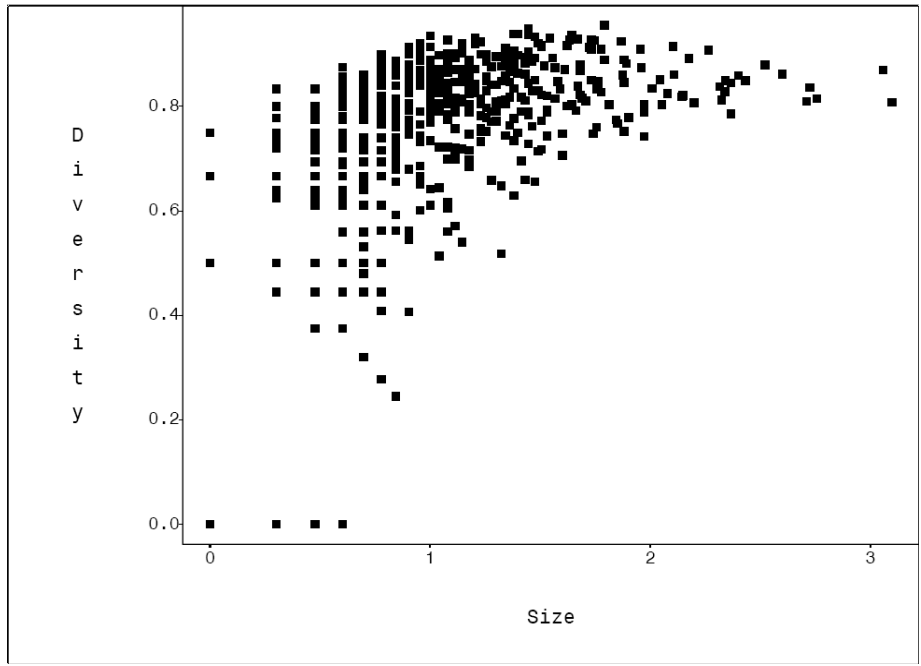

**Figure 3: diversity and hybridization (patents)**

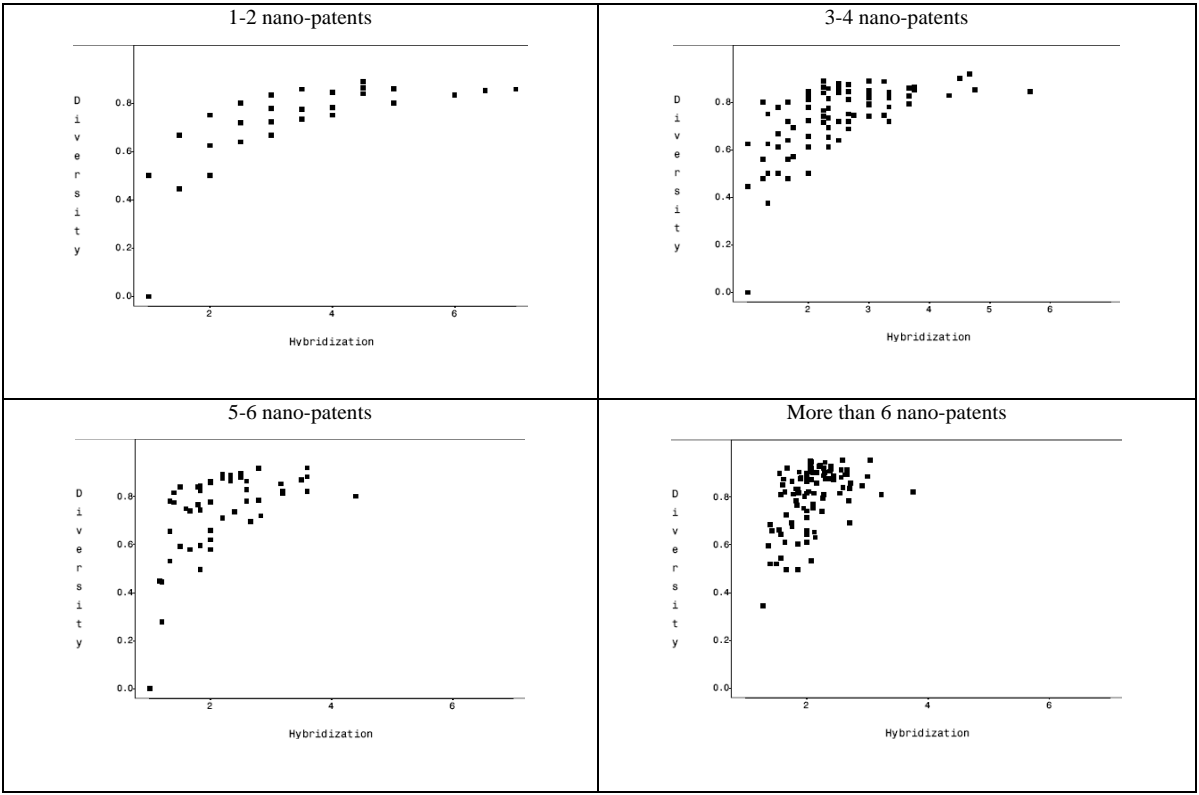

# Figure 4: diversity and hybridization (publications)

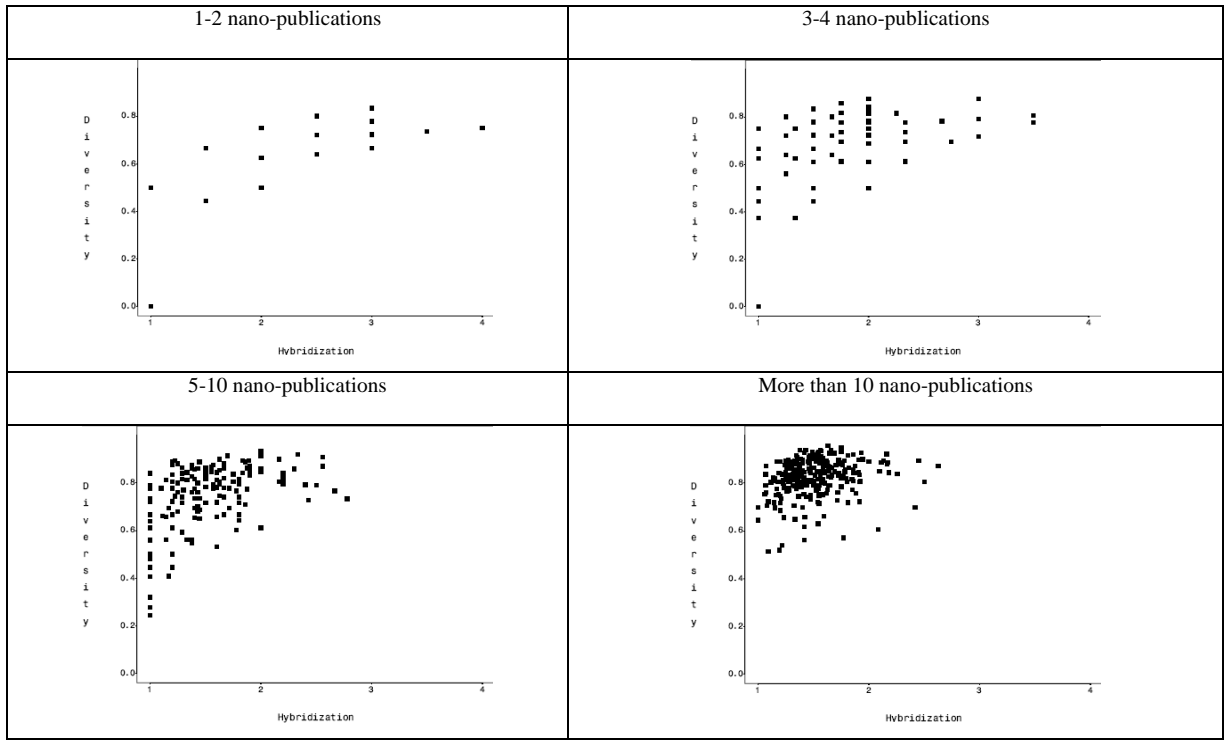